\documentclass[final,3p,times,twocolumn]{elsarticle}

\usepackage{amssymb}
\usepackage{amsmath}
\usepackage{amsthm}
\usepackage{color}

\journal{Journal of Magnetism and Magnetic Materials}

\begin{document}

\begin{frontmatter}



\title{Geometric symmetry and size-dependent skyrmion phase transitions in magnetic nanostructures}


\author[1]{J. Y. Wang} 
\affiliation[1]{organization={School of Materials and Physics, China University of Mining and Technology},
                city={Xuzhou},
                postcode={221116}, 
                country={P. R. China} }
\author[2]{C. X. Zhao\corref{cor1}}  
\ead{cxzhaosxnu@163.com}
\author[1]{Y. F. Duan\corref{cor1}}
\ead{yifeng@cumt.edu.cn}

\affiliation[2]{organization={College of Physics and Information Engineering, Shanxi Normal University},
                postcode={030031}, 
                city={Taiyuan},
                country={P. R. China} }

\author[1]{H. M. Dong\corref{cor1}}

\ead{hmdong@cumt.edu.cn}

\begin{abstract}
We investigate the interplay of geometric symmetry, size, and external magnetic fields in regulating individual skyrmion states within magnetic nanostructures. By analyzing nanodisks, nanosquares, and nanorectangles, we demonstrate that rotational symmetry in nanodisks enables rich topological phase transitions, from ferromagnetic states to skyrmions, skyrmioniums, and multi-states, as their diameter increases. In contrast, square and rectangular structures exhibit suppressed topological complexity due to corner-induced demagnetization effects and reduced symmetries. Under perpendicular magnetic fields, nanodisks show field-driven transitions between skyrmionium and skyrmion states. By leveraging asymmetry, square and rectangular nanostructures stabilize skyrmions over a broader parameter range than nanodisks. These findings highlight geometric symmetry as a critical design parameter for tailoring skyrmion stability and functionality in spintronic applications such as multi-state memory and reconfigurable logic devices.
\end{abstract}


\cortext[cor1]{Corresponding author: C. X. Zhao, Y. F. Duan and H. M. Dong}

\end{frontmatter}



\section{Introduction}
In recent years, the rapid advances in spintronics and topological physics have attracted significant attention to skyrmions, magnetic structures with nontrivial topological properties \cite{fert_magnetic_2017}. A skyrmion is a stable magnetic vortex that can exist at the nanoscale. Its unique topological characteristics offer substantial potential for applications in information storage, processing, and transmission \cite{lisheng2023}. The emergence of skyrmions offers a fresh perspective on traditional magnetic materials and paves the way for the development of innovative spintronic devices. Researchers have made remarkable progress in understanding the mechanisms of skyrmion formation, their stability, and methods for control, thereby establishing a solid foundation for their applications. Although skyrmions possess advantages such as small size, high stability, and low power consumption, making them strong candidates for future information storage and logical operations, effectively manipulating an individual skyrmion in nanostructures remains a significant challenge \cite{casiraghi2019}.

Controlling individual skyrmions depends critically on the design and optimization of nanomagnetic structures, such as racetrack memories that employ nanotracks and nanodisks \cite{ho2019}. Recent research has concentrated on employing geometric design to manage skyrmion states. Very recently, we have shown that in confined nanostructures, new boundary conditions are introduced, resulting in the appearance of tilted skyrmions, due to the presence of the Dzyaloshinskii-Moriya interaction (DMI) \cite{fuSpin2025}. Artificial skyrmion lattices have been created by patterning asymmetric magnetic nanodots with perpendicular magnetic anisotropy (PMA) \cite{Gilbert2015}. Furthermore, neuromorphic weighted-sum operations are implemented using magnetic skyrmions in multilayer tracks \cite{DaCa2025}. Skyrmions in nanoflakes have been observed at room temperature, attributed to compositional gradients across the sample thickness rather than the interfacial DMI \cite{Ni2025}. Skyrmions on a nanowire substrate exhibit attractive interactions that reduce the separation distance between them \cite{Nishitani2024}. Research has demonstrated that geometrically confined magnetic nanorings can be used to construct skyrmion devices with novel functionalities \cite{Kechrakos2023}. Twisted skyrmions are generated in multilayered nanostructures due to the competition between DMI and magnetic dipole-dipole interactions \cite{boTailor2021}. Additionally, the helicity of vortex-like skyrmions can be adjusted by modifying material parameters and the geometry of soft ferromagnetic nanodots and films \cite{verba2020}. Skyrmions can also take the shape of a circular truncated cone by disrupting the helical stripes on the left and right sides of the nanotube. These studies illustrate that nanostructure skyrmion states possess unique properties and applications, attracting increasing attention and research. However, there remains a lack of in-depth research on how different magnetic nanostructures tune skyrmion states \cite{liuRoom2025}, particularly regarding their shapes and sizes. 

Individual skyrmions can be effectively controlled and manipulated using geometrical magnetic nanostructures. Research has shown that sub-100-nm Néel-textured skyrmions can be stabilized at room temperature, without an external magnetic field, when confined in nanodots across a wide range of magnetic and geometric parameters \cite{ho2019}. These geometrically confined skyrmions can vary in size and shape, exhibiting a range of sizes and ellipticities in a nanostripe not present in thin films or bulk materials \cite{jin2017}. Additionally, selectively grown nanowires of cubic FeGe can support stable skyrmions \cite{stolt2017}. In fully confined geometries, the sharp edges can induce natural oscillations, leading to dynamic resonances in the skyrmions \cite{navau2016}. The orientation of a square skyrmion lattice can be manipulated by adjusting the edges of the nanostructures and by considering the mismatched symmetries between the island shape and the skyrmion lattice \cite{hageme2016}. Pepper et al. demonstrated that square and triangular nanostructures can host skyrmion ground states while suppressing higher‑order textures. However, a systematic comparison of nanodisks, squares, and rectangles under simultaneous size and field variations, together with an energy‑density‑based analysis and the field‑driven skyrmionium-skyrmion transition, remains missing. The present work aims to fill this gap \cite{perpper2018}. Furthermore, a series of skyrmion cluster states with varying geometric configurations have been identified, along with field-driven cascading phase transitions occurring at temperatures well below the magnetic transition temperature \cite{zhaoDirect2016a}. Our research has uncovered novel non-trivial topological structures, including $3\pi$-skyrmions and flower-like and windmill-like skyrmions in nanodisks, which warrant further investigation \cite{dong_jap}. Notably, the mechanisms by which magnetic nanostructures and their sizes influence skyrmion states remain largely unknown \cite{du2022}. How these structures and sizes interact with magnetic parameters remains poorly understood \cite{reich2022}. We are conducting a systematic theoretical study of micromagnetism to address these gaps. We aim to gain a deeper insight into how magnetic nanostructures and their parameters regulate the topological states of skyrmions in different nanostructures, particularly focusing on the combined effects of shape, size, and magnetic parameters.

\section{Theoretical Approaches}

This work presents a comprehensive exploration of research theories and models that investigate the intricate micromagnetic nanotopology associated with individual spin topological states. In magnetic nanostructures, boundary effects play a crucial and multifaceted role in shaping the overall magnetic behavior and characteristics \cite{luSize2026}. These boundary effects, arising from the unique spatial confinement and surface interactions within the nanostructures, can significantly influence the emergence and stability of various spin-topological states \cite{fuSpin2025}. The shape of the nanostructure, including its geometric configuration and associated demagnetization factors, also serves as a critical regulator of the resulting topological magnetic structure. 

Furthermore, the comprehensive investigation undertaken in this study can serve as a springboard for nanomagnetic research, enabling the exploration of more advanced and complex magnetic nanostructures. The insights gained from this work can contribute to the development of innovative technologies that leverage the unique properties of spin topological states, opening new frontiers in fields such as spintronics, magnetic data storage, and quantum computing. In a magnetic nanosystem, the Gibbs free energy is a crucial physical quantity that describes the system's state. Various energy components of the system can be constructed by defining the continuous vector field of magnetization $\boldsymbol{M}$. 

The total Gibbs free energy $E_{\text{tot}}$ of a magnetic nanosystem described by the continuous magnetization vector field $\boldsymbol{M}$ is expressed as a sum of competing contributions~\cite{dong2023,muller2019}:
\begin{equation}
E_{\text{tot}} = E_{\text{ex}} + E_{\text{DMI}} + E_{\text{ani}} + E_{\text{Zeem}} + E_{\text{deg}},
\label{eq:total}
\end{equation}
where $E_{\text{ex}}$ is the symmetric exchange energy, $E_{\text{DMI}}$ the antisymmetric interfacial DMI energy, $E_{\text{ani}}$ the magnetic anisotropy energy, $E_{\text{Zeem}}$ the Zeeman energy, and $E_{\text{deg}}$ the demagnetization (magnetostatic) energy.

At equilibrium, the normalized magnetization $\boldsymbol{m}$ satisfies the Brown condition by the saturation magnetization $M_s$ satisfies the Brown condition
\begin{equation}
\boldsymbol{m} \times \boldsymbol{H}_{\text{eff}} = 0, \quad 
\boldsymbol{H}_{\text{eff}} = -\frac{1}{\mu_0 M_s} \frac{\delta E_{\text{tot}}}{\delta \boldsymbol{m}},
\label{brown}
\end{equation}
where $\boldsymbol{H}_{\text{eff}}$ is the total effective field. $\mu_0$ is the magnetic permeability in a vacuum. The dynamical evolution is governed by the Landau-Lifshitz-Gilbert (LLG) equation~\cite{Joos2023}
\begin{equation}
\frac{\partial \boldsymbol{m}}{\partial t} = -\gamma \boldsymbol{m} \times \boldsymbol{H}_{\text{eff}} - \alpha \boldsymbol{m} \times \left( \boldsymbol{m} \times \boldsymbol{H}_{\text{eff}} \right),
\label{llg}
\end{equation}
with $\gamma$ the gyromagnetic ratio and $\alpha$ the Gilbert damping constant.

We solve the LLG equation numerically using the finite-difference GPU-accelerated package Mumax3 \cite{mumax3}. The nanostructures (disks, squares, rectangles) have a uniform thickness of $1$~nm and are discretized with a regular mesh of $1 \times 1 \times 1$~nm$^3$. Convergence with respect to mesh refinement was verified by repeating selected simulations with a $0.5 \times 0.5 \times 1$~nm$^3$ cell size; the resulting phase boundaries shifted by less than $1$~nm. Free surface boundary conditions are applied, together with the special DMI boundary conditions required at the edges of confined structures~\cite{rohart2013}. The initial magnetization configuration was random. The ground state was obtained by energy minimization using the conjugate‑gradient‑based Relax() routine. Convergence was assumed when the maximum torque dropped below $10^{-5}\ \text{T}$. We have then allowed sufficient relaxation and testing time to ensure that the system has reached its ground state.

Based on our previous work~\cite{dong_jap,dong_2023}, the following material parameters are adopted: $M_s = 914$~kA/m, the symmetric exchange stiffness $A = 11.2$~pJ/m, the interfical DMI constant $D = 4.1$~mJ/m$^2$, the PMA constant $K_u = 6$~MJ/m$^3$, and $\alpha = 0.1$. All simulations were carried out at $T = 4.2~\text{K}$ without a thermal fluctuation term, because the aim of this work is to elucidate the ground‑state topological phase diagrams governed by geometric symmetry and size. Although thermal activation at room temperature may shift phase boundaries and reduce the lifetime of metastable states such as skyrmionium, the low‑temperature results capture the intrinsic energy landscape. The influence of finite temperature is beyond the scope of the present work and will be addressed in future studies. Although simultaneously achieving high $K_u$ and $D$ values is challenging, recent multilayer engineering has approached similar parameter ranges; the qualitative trends reported here are robust for lower, experimentally accessible parameters as well. Simultaneously achieving $K_u = 6\ \text{MJ/m}^3$ and $D = 4.1\ \text{mJ/m}^2$ is demanding, although recent experiments on optimized multilayers have demonstrated $K_u > 4\ \text{MJ/m}^3$ and $D = 2 \sim 3\ \text{mJ/m}^2$ \cite{liuStrong2017,zhu2024a}. The parameter set used here should therefore be viewed as a theoretical limit; the qualitative dependence of the topological phase boundaries on geometry is robust also for smaller $K_u$ and $D$ values.

\section{Results and Discussions}

\begin{figure*}[htb]
\centering
\includegraphics[width=0.60\textwidth,height=0.60\textheight,angle= 0]{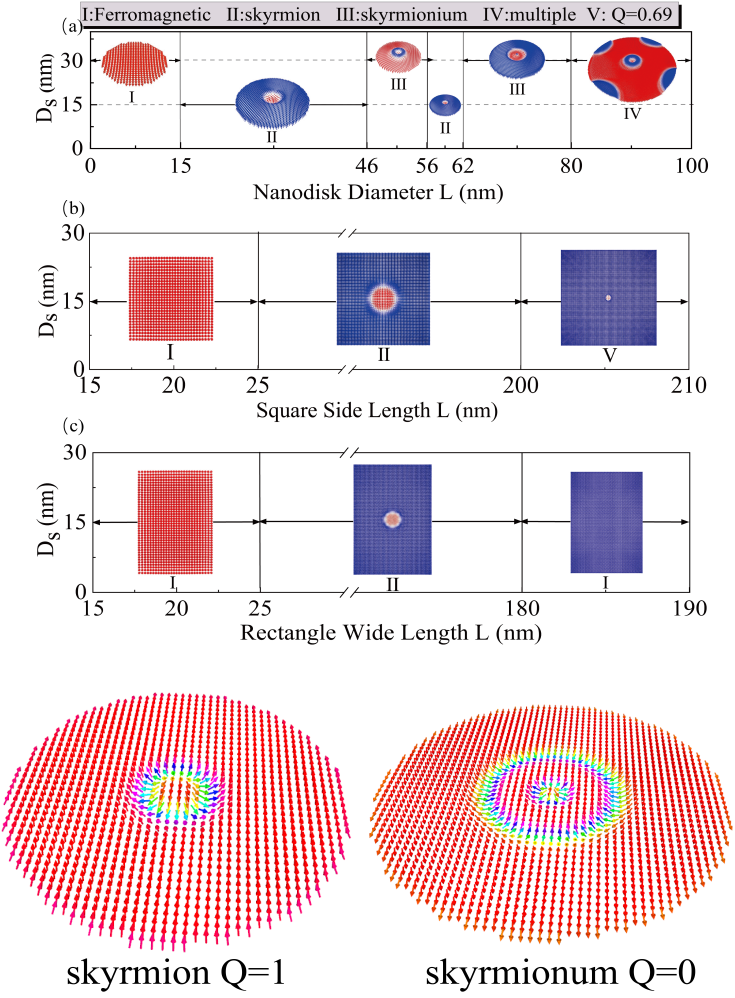}
\caption{The size-dependent magnetic topological structures with the diameter $D_s$ of (a) a nanodisk, (b) a square, and (c) a rectangle. The inserts show schematics of magnetic topological structures with topological charge $Q$. The typical spin configurations, skyrmion and skyrmionum, are shown, respectively.} \label{fig1}
\end{figure*}
We present the simulation results illustrating the size $L$-dependent magnetic topological structures with the spin structure diameter $D_s$ of (a) a single-layer nanodisk, (b) a single-layer square, and (c) a single-layer rectangle in Fig. \ref{fig1}, respectively. At the transition points, we conducted numerous simulations to gather precise data. The topological states in the nanodisk exhibit remarkable diversity, as shown in Fig. \ref{fig1}(a). In the nanodisk, skyrmions (II) and skyrmioniums (III) can transform into each other when the diameter falls between 40 nm and 80 nm. This phase further evolves into the skyrmionium state (III) and a multi-state (IV) within a diameter range of $70-100$ nm. This multi-level phase transition is attributed to the enhancement of geometric symmetry (rotational symmetry $C_N$) on DMI and to the tuning of magnetic anisotropy. The multi-state appearance indicates that the disk's continuous symmetry reduces the distortion of the magnetic moments, allowing for the stable existence of high-order topological defects, such as multi-skyrmions. In contrast to the nanodisk, we observe topological suppression phenomena in the square and rectangle, as shown in Figs. \ref{fig1}(b) and (c). The square and rectangle exhibit nontrivial spin-topological states only at specific sizes. They demonstrate only simple topological configurations, such as the single skyrmion state. The reduction of symmetry and the strong local demagnetizing field arising from sharp corners disrupt long-range magnetic order, suppressing the formation of complex topological states. The three structures have varying demagnetization factors $\mathcal{N}$, as previously mentioned \cite{Bahl2021}. The aspect ratio of $1.5:1$ in the rectangular structure enhances shape anisotropy. This causes the magnetic moments to preferentially align along the long axis, thereby weakening the DMI contribution, which depends on the coupling between the out-of-plane magnetic moment components.

Additionally, the energy density within the nanodisk increases with its diameter, as shown in Fig. \ref{fig1}. The geometric symmetry tunes the competition between DMI and exchange energy. For small sizes ($L < 40$ nm), the exchange energy is dominant, resulting in the ferromagnetic state (I) having the lowest energy. As the size becomes moderate, about $50 < L < 70$ nm, the DMI energy increases, allowing the skyrmion state (II) to lower the overall energy through topological protection. For larger sizes ($L > 80$ nm), the demagnetization energy, which is proportional to the volume, increases, prompting the system to reach energy balance by splitting the topological charge, and can lead to states such as skyrmioniums. At the corners of square and rectangular shapes, the demagnetizing field significantly increases the demagnetization energy, thereby lowering the system's energy by forming single topological states. We note that the present zero‑temperature phase diagrams provide a reference frame. The actual spin structure must account for the effects of temperature at room temperature, and the spin topological structure must exist below the Curie temperature.

For rectangular structures, the size threshold, about 180 nm, relates to the transition from a single-skyrmion to a ferromagnetic state, primarily driven by demagnetization energy, as shown in Fig. \ref{fig1}. However, due to geometric asymmetries, rectangular structures cannot support high-order topological states. Our research demonstrates that nanodisks, with their diverse range of phase transitions, are excellent candidates for applications such as multi-state memory or reconfigurable spin devices. In contrast, rectangular structures offer more topological stability. Still, their limited variety of topological states makes them better suited for magnetic logic units that require high resistance to interference, such as skyrmion racetrack memories \cite{Fert2013}. Notably, a size deviation of just 5-10 nm can significantly alter the topological state, underscoring the importance of precision in nanomanufacturing for optimal device performance. While the disk phase diagrams were partly reported in our earlier work \cite{dong_jap, dong2023}, the present study provides the missing comparative perspective by including squares and rectangles, thereby revealing that symmetry reduction suppresses topological diversity but, counterintuitively, enhances the robustness of the elementary skyrmion state. This finding provides a quantitative theoretical understanding for designing magnetic devices based on the regulation of shape and size. The fractional topological charge, such as $Q = 0.69$, is observed in nanoscale magnetic structures under confined conditions. Although theoretically, a topological invariant $Q$ is defined on an infinite two-dimensional space. In finite-sized magnets, the presence of incomplete topological magnetic structures in these magnets leads to a collective fractional topological charge.

\begin{figure}[htb]
\centering
\includegraphics[width=0.49\textwidth,height=0.25\textheight,angle= 0]{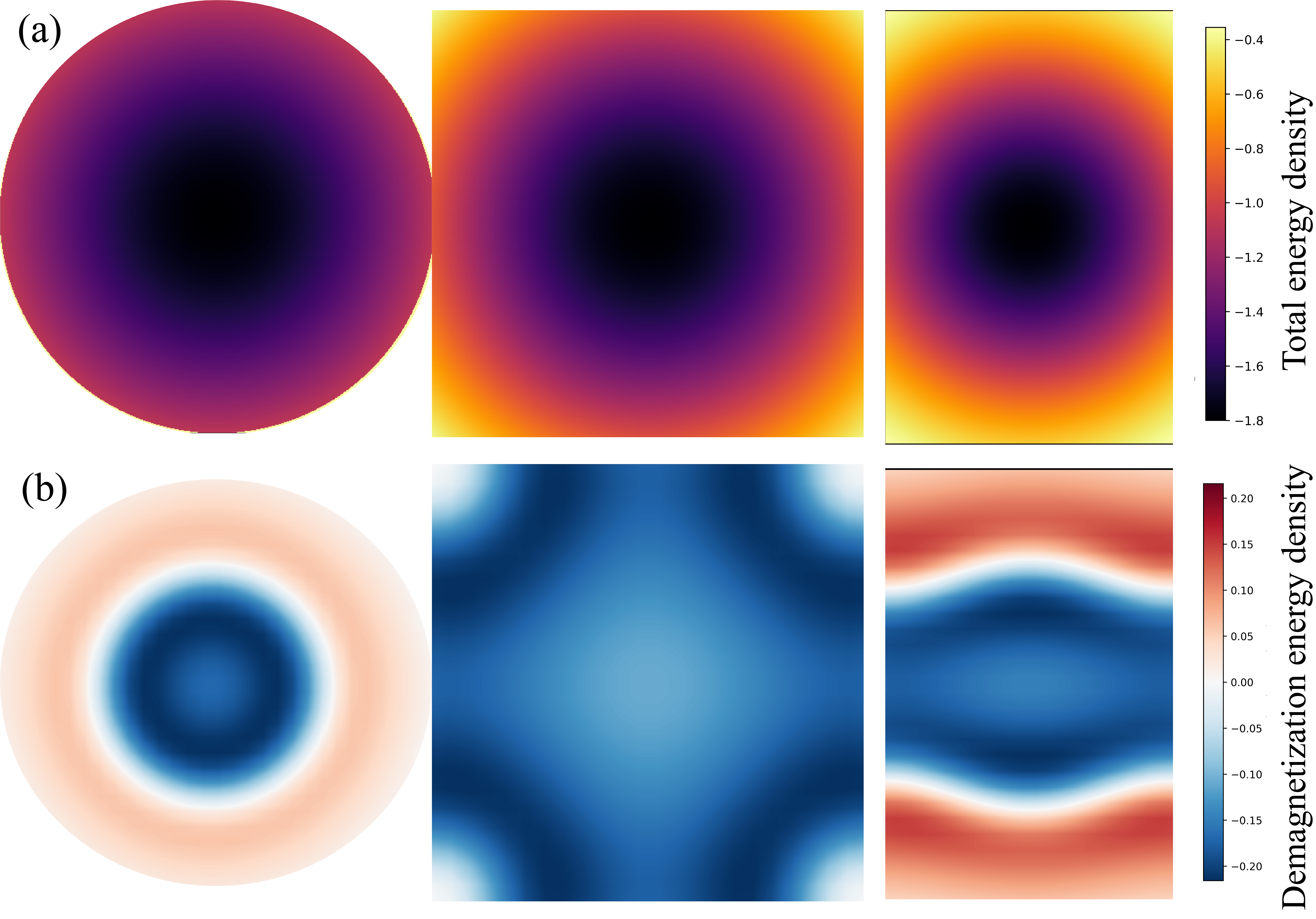}
\caption{The distributions of magnetic energy density (a) and the demagnetization energy density (b) in a nanodisk with a diameter of 50 nm, a square with $50\times50$ nm, and a rectangle with $50\times75$ nm without the magnetic field ($B=0$).} \label{fig2}
\end{figure}

We present the distributions of magnetic energy density for various magnetic nanostructures to aid the interpretation of our theoretical results, as shown in Fig. \ref{fig2}(a). In the nanodisk case, the magnetic energy density distribution exhibits a high degree of symmetry. The region of energy concentration is annularly arranged around the skyrmion core, with a relatively gentle energy gradient at the edge, indicating a strong topological protection effect. For the square structure, the energy distribution is significantly constrained by the geometric boundaries. Multiple energy concentration points are observed around the skyrmion core, with the magnetic energy density increasing notably at the corners. In the rectangular structure, the distribution of magnetic energy density demonstrates considerable anisotropy. The energy gradient along the long axis is steeper, while the energy diffusion range along the short axis is broader. This reflects the stretching effect of the rectangle's shape on the magnetic structure. The influence of right-angled boundaries is evident in both the square and rectangular configurations, particularly in the rectangle. An increase in the aspect ratio of the rectangle enhances the magnetic anisotropy. The differences in magnetic energy density distribution among the various geometric structures are primarily driven by symmetry breaking and boundary effects. Moreover, the demagnetization energy density maps [Fig. \ref{fig2}(b)] reveal intense, localized peaks at the corners of the square and rectangle, which are absent in the nanodisk. These corner fields act as pinning barriers that fragment the chiral texture, explaining the topological suppression observed in Figs. \ref{fig1}(b) and 1(c).

\begin{figure}[htb]
\centering
\includegraphics[width=0.49\textwidth,height=0.30\textheight,angle= 0]{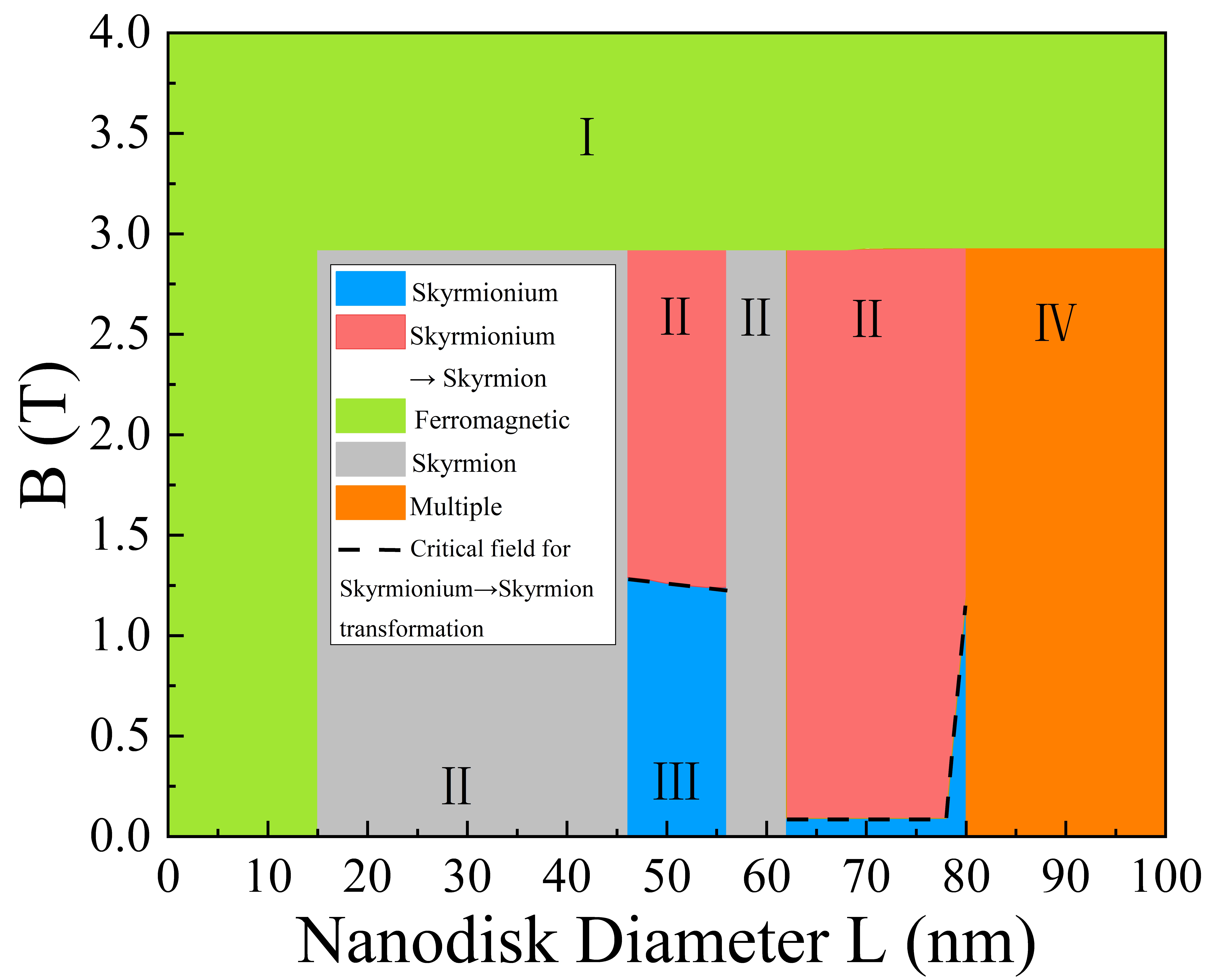}
\caption{The topological phase diagram of nanodisks as a function of a perpendicular magnetic field $B$ and its diameter $L$.} \label{fig3}
\end{figure}
Fig. \ref{fig3} illustrates the topological phase diagram in a nanodisk as a function of the perpendicular magnetic field $B$ and its diameter $L$. This diagram illustrates the reconstruction of topological states induced by the magnetic field and highlights their key characteristics. The phase diagram features four distinct types of topological states: (i) Ferromagnetic state (I) prevails in regions with high magnetic fields ($B > 2.9 \, \text{T}$) and small diameters ($L < 15 \, \text{nm}$). In this state, the magnetic moments are fully aligned with the magnetic field, thereby suppressing topological defects. (ii) The transition from skyrmioniums to skyrmions (red) induced by magnetic fields occurs in regions of moderate magnetic fields and nanosizes ($ 46 < L < 80 \, \text{nm}$). Skyrmioniums (III) exist only within the range of low magnetic fields and very small ranges of diameters. The circular distribution of its magnetic moment provides topological protection against external field perturbations. (iii) Skyrmion State (II) dominating areas with moderate sizes ($50 \, \text{nm} < L < 80 \, \text{nm}$), this state is characterized by its single-core structure, which is stabilized by the magnetic field compressing the topological charge density. (iv) Multi-State (IV) is observed only in regions of large size ($L > 80 \, \text{nm}$) and low magnetic field ($B < 2.0 \, \text{T}$). 

An important discovery highlighted in this study is the topological phase transition driven by a critical magnetic field $B_c$ (approximately 1.5 to 1.6 T), which facilitates the transition from skyrmionium to skyrmion. When the magnetic field exceeds $B_c$, the circular structure of skyrmionium collapses into a single-core skyrmion. This occurs because the magnetic energy (Zeeman energy) surpasses the energy barrier provided by topological protection, leading to a sudden change in the topological charge from $Q = 0$ (skyrmionium) to $Q = 1$ (skyrmion). In a perpendicular magnetic field, magnetic moments tend to align with the field. As the magnetic field strength increases, the radius of the circular magnetic moment distribution of skyrmionium decreases until it becomes unstable at the critical field. The core radius of the skyrmion state also diminishes with increasing $B$, ultimately becoming completely suppressed, transitioning into a ferromagnetic state. Furthermore, the critical magnetic field changes abruptly with increasing diameter, primarily due to a discontinuous change in the topological charge. When the system size is small (about $L < 50$ nm), the exchange energy ($E_{\text{ex}} \propto 1/L^2$) increases significantly, which suppresses the formation of topological defects. On the other hand, when the size is larger (about $L > 80$ nm), the competition between the demagnetization energy and the DMI results in a multi-state configuration. In this case, the magnetic field selects the lowest-energy state by adjusting the energy balance among competing interactions. It is noteworthy that Sarmiento et al. recently studied this same transition in nanodisks at finite temperature, extracting energy barriers and skyrmionium lifetimes. Our zero‑temperature critical field $B_c$ is consistent with the onset of instability found in that work. A full thermal‑activation analysis is beyond the scope of the present study, but will be pursued in the future \cite{Sarmiento2024}.

In the absence of a magnetic field, an increase in diameter initiates a sequence of states, for instance, from ferromagnetic state to skyrmion to skyrmionium, to multi-state. This transition is driven by the balance between geometric symmetry and demagnetization energy. Zeeman energy is introduced as an additional contribution in the presence of a magnetic field. This facilitates the reconstruction of topological states, such as the interconversion between skyrmionium and skyrmion, and enables the reversible switching of these states through the $B-L$ phase diagram \cite{dong_jap}. This approach expands the functional dimensions of the nanodisks. By using the $B-L$ phase diagram to design a multi-state memory, we can achieve information encoding that surpasses binary limits by adjusting the magnetic field and structural size. The critical field required for the skyrmionium to skyrmion transition is highly sensitive to size and can serve as a nanoscale magnetic field sensor. The study highlights the significant impact of the interplay between the perpendicular magnetic field and nanodisk size on the evolution of topological states. The results clarify the mechanism of the topological phase transition driven by the critical field and provide essential theoretical support for the design and optimization of multifunctional magnetic nanodevices \cite{Xing2025}.

\begin{figure}[htb]
\centering
\includegraphics[width=0.49\textwidth,height=0.30\textheight,angle= 0]{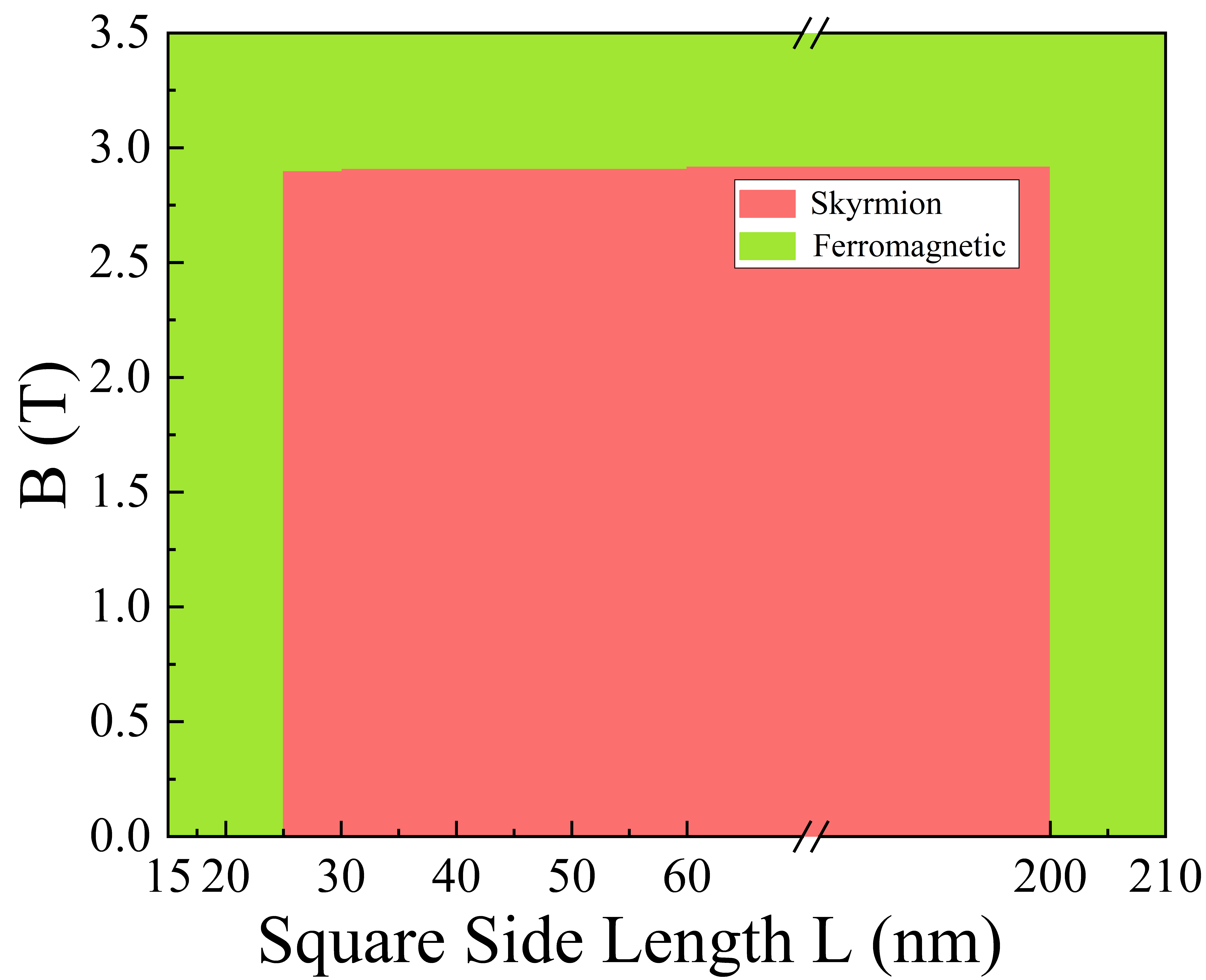}
\caption{The topological structure phase diagram of the magnetic square as a function of the perpendicular magnetic field $B$ and the side length $L$.} \label{fig4}
\end{figure}
The topological phase diagram of a magnetic square depending on perpendicular magnetic fields $B$ and the side length $L$, is illustrated in Fig. \ref{fig4}. The introduction of a perpendicular magnetic field allows a single skyrmion state (the red area) to stably exist over a wider range for $25 < L < 200$ nm and $B < 2.8$ T. This phenomenon indicates that the magnetic field plays a crucial role in regulating the energy competition between different interactions in nanostructures. The magnetic field aligns the magnetic moments partially along its direction, thereby enhancing the gradient distribution of the perpendicular components of the magnetic moment. Additionally, as the demagnetization energy $E_{\text{deg}}$ decreases, the magnetic field effectively reduces the out-of-plane demagnetizing field. This mitigates local demagnetization effects typically observed at the corners of the square, thereby creating a more uniform magnetic environment conducive to the stability of topological states.

In comparison to the behavior of the nanodisk under a magnetic field, the nanodisk can stabilize skyrmioniums (topological charge $Q = 0$) and multiple states. In contrast, the square structure only supports a single skyrmion state (topological charge $Q = 1$). This difference arises because the rotational symmetry of the nanodisk facilitates circular or helical magnetic moment distributions, which allow for the existence of skyrmioniums \cite{dong_jap,yang2023a}. In contrast, the right-angled corners of the square introduce a strong anisotropic demagnetizing field, disrupting the continuity of the long-range magnetic order and suppressing the formation of higher-order topological defects. The spatial modulation of the DMI interaction induces abrupt changes in the direction of the DMI vector along the square's boundary, thereby hindering the expansion of the chiral magnetic texture and limiting the accumulation of the topological charge number. In the $B-L$ phase diagram of the nanodisks, the region representing the skyrmion state expands non-linearly, while the phase boundary for the square structure is approximately linear. This reflects the differential influence of shape on the balance of energy competition. The competition between demagnetization energy ($\propto L^3$) and exchange energy ($\propto 1/L^2$) primarily drives the phase transition in the nanodisks. Conversely, in the square shape, the localization of the demagnetizing field at the corners makes the energy competition more sensitive to the direct modulation of the DMI energy by the magnetic field. 

Compared to the nanodisk, the square, with only skyrmion and ferromagnetic states, has a more limited set of topological states, making it more suitable for magnetic storage units that perform binary operations. On the other hand, the nanodisk's multi-state characteristics are more appropriate for high-density information encoding. This phase diagram illustrates the crucial regulatory role of the perpendicular magnetic field on the topological states of the square. By balancing Zeeman energy, DMI energy, and demagnetizing energy, the magnetic field can not only induce the formation of the skyrmion state but also affect its stability. The comparison with the nanodisk emphasizes the significant impact of geometric symmetry on the diversity of topological states. This finding provides a key theoretical foundation for designing magnetic devices that leverage the synergistic regulation of shape and magnetic field, offering valuable insights regarding the controllability of topological states, size compatibility, and resistance to interference (see also Fig. \ref{fig2}).

\begin{figure}[htb]
\centering
\includegraphics[width=0.49\textwidth,height=0.30\textheight,angle= 0]{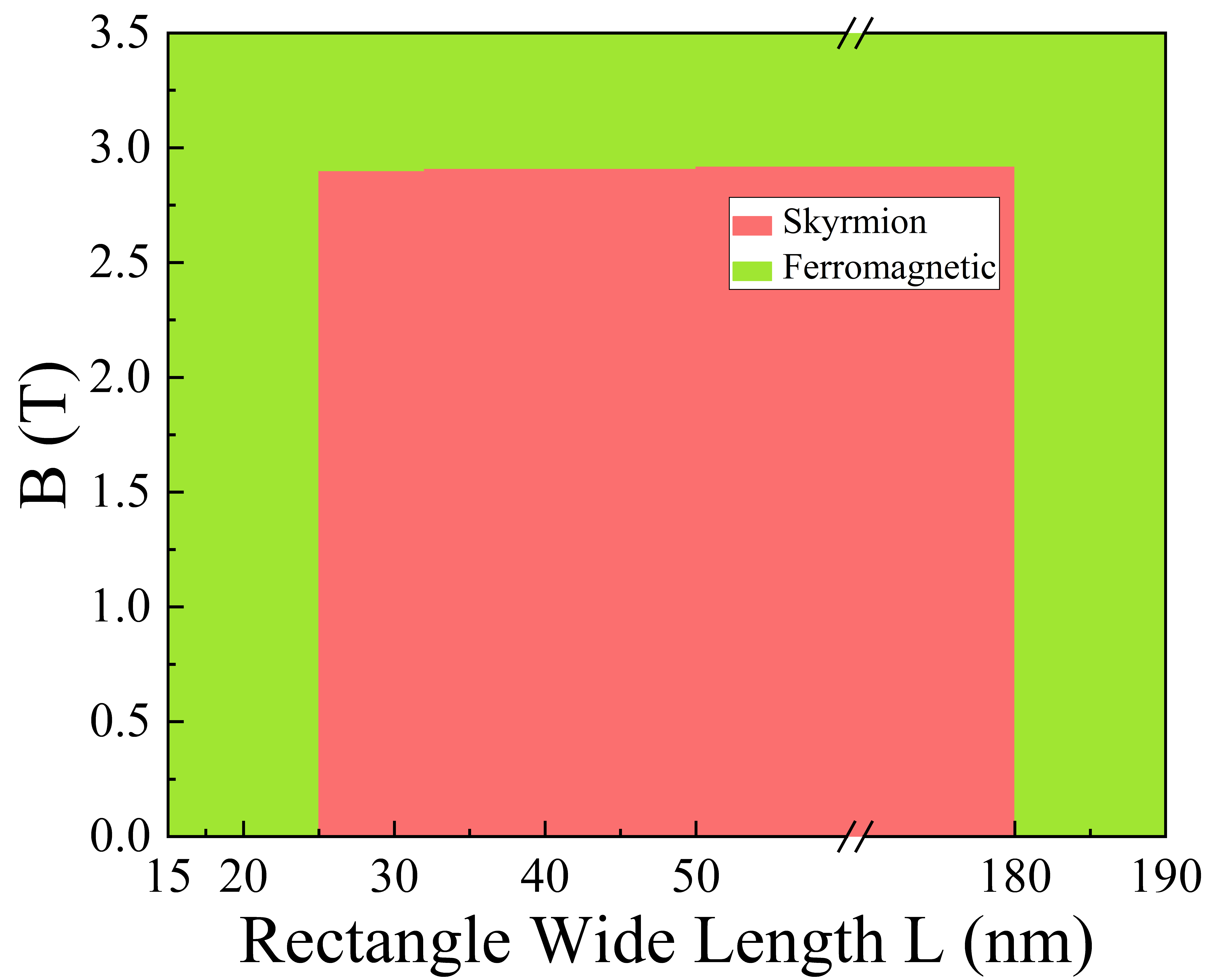}
\caption{The topological structure phase diagram of the magnetic rectangle (with the aspect ratio 1.5:1) under a magnetic field $B$ and its width $L$.} \label{fig5}
\end{figure}

Figure~\ref{fig5} presents the topological phase diagram of a magnetic rectangle 
(aspect ratio $1.5:1$) as a function of its width $L$ and the perpendicular magnetic 
field $B$. The single-skyrmion state remains stable for $25 < L < 180$~nm and 
$B \lesssim 2.92$~T. Compared with the square, the stability window is extended toward 
higher magnetic fields, although the upper size limit is slightly reduced. This behavior 
originates from the shape anisotropy of the rectangle. The smaller demagnetization 
factor along the long axis reduces the in-plane shape anisotropy along that direction, 
weakening the local tilting of magnetization and helping to preserve a more uniform 
perpendicular magnetization component. The ferromagnetic state prevails when 
$L < 25$~nm, $L > 180$~nm, or $B \gtrsim 2.92$~T. In the small-size regime, the 
exchange energy dominates, suppressing non-collinear textures, while under strong 
fields, the Zeeman energy overcomes the DMI, forcing the moments to align 
collinearly. To further isolate the role of asymmetry, we also simulated rectangles 
with a fixed long side and a variable short side; the results show a similarly 
enlarged skyrmion region, confirming that the asymmetric shape alone broadens the 
stability window relative to the square.

The asymmetric geometry of the rectangle reduces the localization of the 
demagnetizing field at the corners (see Fig.~\ref{fig2}), and the elongation along the 
long axis smooths the spatial variation of the DMI vector, thereby lowering the local 
DMI energy gradient compared to the square. In the medium-size range 
($25~\mathrm{nm} < L < 180~\mathrm{nm}$), an energy balance is established among the 
exchange, DMI, and Zeeman terms, allowing the skyrmion to remain a low-energy 
state through topological protection. The combination of a perpendicular magnetic 
field and the asymmetric shape synergistically broadens the skyrmion-stable domain 
by adjusting the competition between demagnetization and DMI energies. These 
findings offer practical guidance for designing robust spintronic devices, particularly 
for high-density storage and programmable logic gates \cite{Dieny2020}.

\begin{figure}[htb]
\centering
\includegraphics[width=0.49\textwidth,height=0.30\textheight,angle= 0]{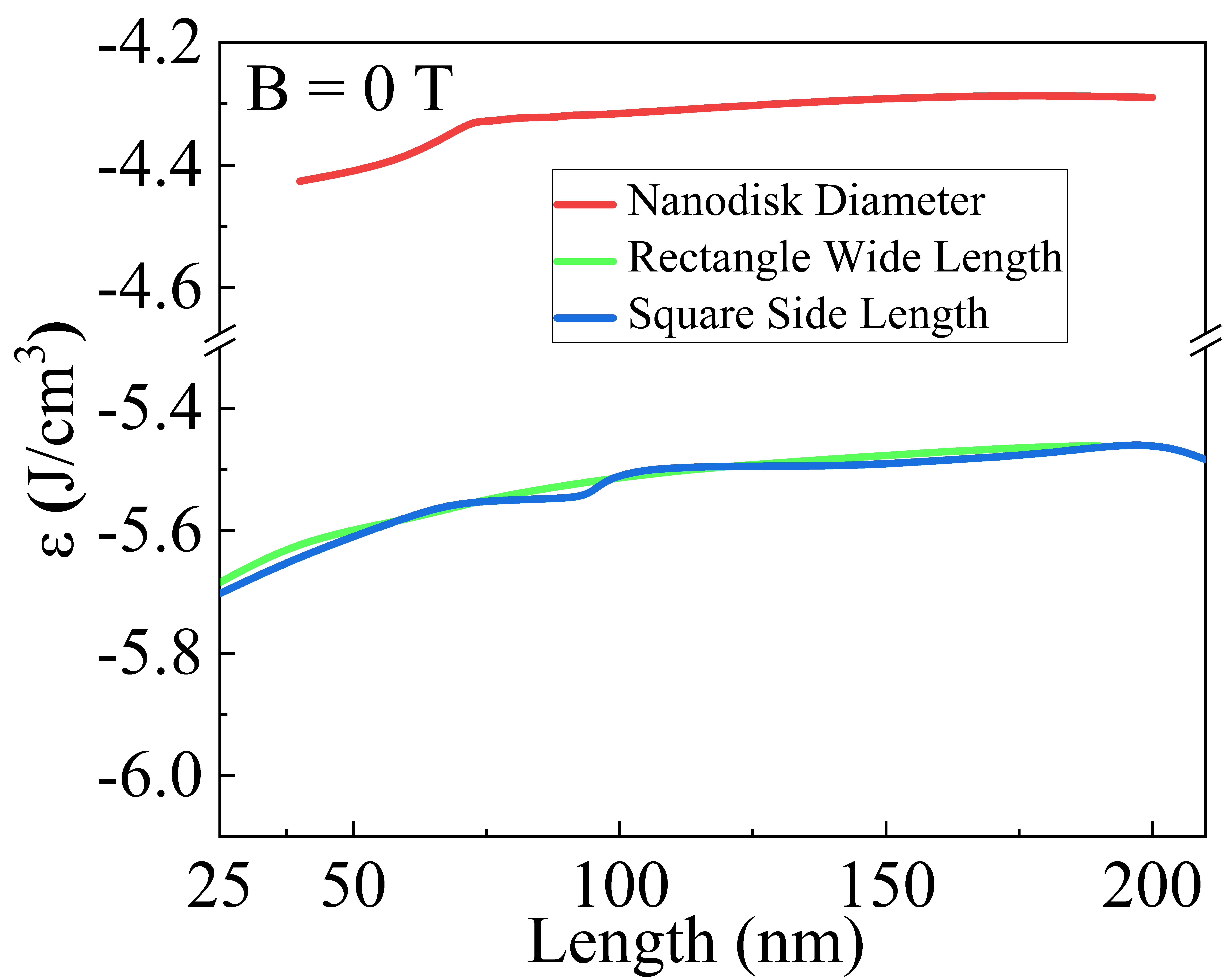}
\caption{The variation of the total magnetic energy density $\varepsilon_{tot}$ in different magnetic nanostructures with size.} \label{fig6}
\end{figure}
Fig. \ref{fig6} illustrates the variation in total magnetic energy density $\varepsilon_{tot}$ across different magnetic nanostructures as a function of size. The magnetic energy density of the nanodisk is significantly higher than that of the square and rectangular structures. This is primarily due to the nanodisk's continuous rotational symmetry, which enhances the total magnetic energy density through several mechanisms. The shape anisotropy of the disk compels the magnetic moments to form a closed topological structure, for example, a vortex state or a skyrmionium state, within the plane. This results in a larger proportion of magnetic moments deviating from the easy magnetization direction. Consequently, the contribution of the anisotropy energy $\varepsilon_{\text{ani}}$ is significantly higher. In Figure \ref{fig6}, the proportion of anisotropy energy for the nanodisk in the size range of 25 to 200 nm consistently surpasses that of the square and rectangular shapes. The disk's symmetry allows the DMI vector to be continuously distributed along the radial or circumferential directions (see Fig. \ref{fig2}). This promotes a gradient arrangement of non-collinear magnetic moments, such as the helical structure of a skyrmion, maximizing the contribution of DMI energy.

In contrast, the right-angled boundaries of the square and rectangular shapes lead to suppressing the accumulation of DMI energy, as shown in Fig. \ref{fig2}. The complex topological structures present in the disk (such as multiple skyrmions or skyrmioniums) require increased exchange energy to maintain the continuous variations of adjacent magnetic moments, depending on the square of the magnetic moment gradient. Notably, in the medium size range (50-100 nm), the competition between exchange energy and DMI energy results in a further increase in the total energy density in Fig. \ref{fig6}. Previous simulations have demonstrated that, in the absence of a magnetic field, a nanodisk undergoes a sequence of phase transitions: from a ferromagnetic state to a skyrmion state, then to a skyrmionium state, and finally to a multistate as its size increases. The stability of these high-order topological states relies on maintaining a high energy density. The circular magnetic moment distribution in the skyrmionium state ($Q = 0$) demands greater DMI energy and exchange energy to sustain its gradient structure. The coexistence of multiple skyrmions leads to a superposition of domain-wall and exchange energies, further increasing the total energy density. This high energy density grants the topological state enhanced resistance to thermal fluctuations and defect pinning, effectively creating an energy barrier that helps maintain a steady state. In contrast, due to geometric constraints, square and rectangular configurations can only form simple topological states, such as a single skyrmion or a vortex state. Their energy density is significantly lower because it is primarily influenced by demagnetization energy. The disk reaches an energy (approximately $-4.3$ J/cm$^3$) due to strong exchange and anisotropy energies, primarily in the ferromagnetic state. In comparison, the energy density of the square and rectangle is relatively low (around $-5.6$ J/cm$^3$) because the suppression of demagnetization energy has not yet taken effect.

The disk mitigates the disadvantages of demagnetization energy through its multistate configuration, thereby maintaining a higher energy density than the square and rectangular configurations. This reflects the capability of topological protection to uphold high energy density. The elevated magnetic energy density of the nanodisk arises from the synergistic enhancement of its geometric symmetry across multiple energy terms (anisotropy, DMI, and exchange), coupled with a positive feedback between the stability of the topological state and the energy density. This characteristic endows the nanodisk with high stability and controllability in spintronic devices. Conversely, the square and rectangle, affected by multi-domain splitting due to demagnetization energy, are more suitable for low-energy-consumption magnetic storage applications. This discovery provides a crucial theoretical foundation for the energy optimization and functional design of nanomagnetic devices.

\begin{figure}[htb]
\centering
\includegraphics[width=0.49\textwidth,height=0.30\textheight,angle= 0]{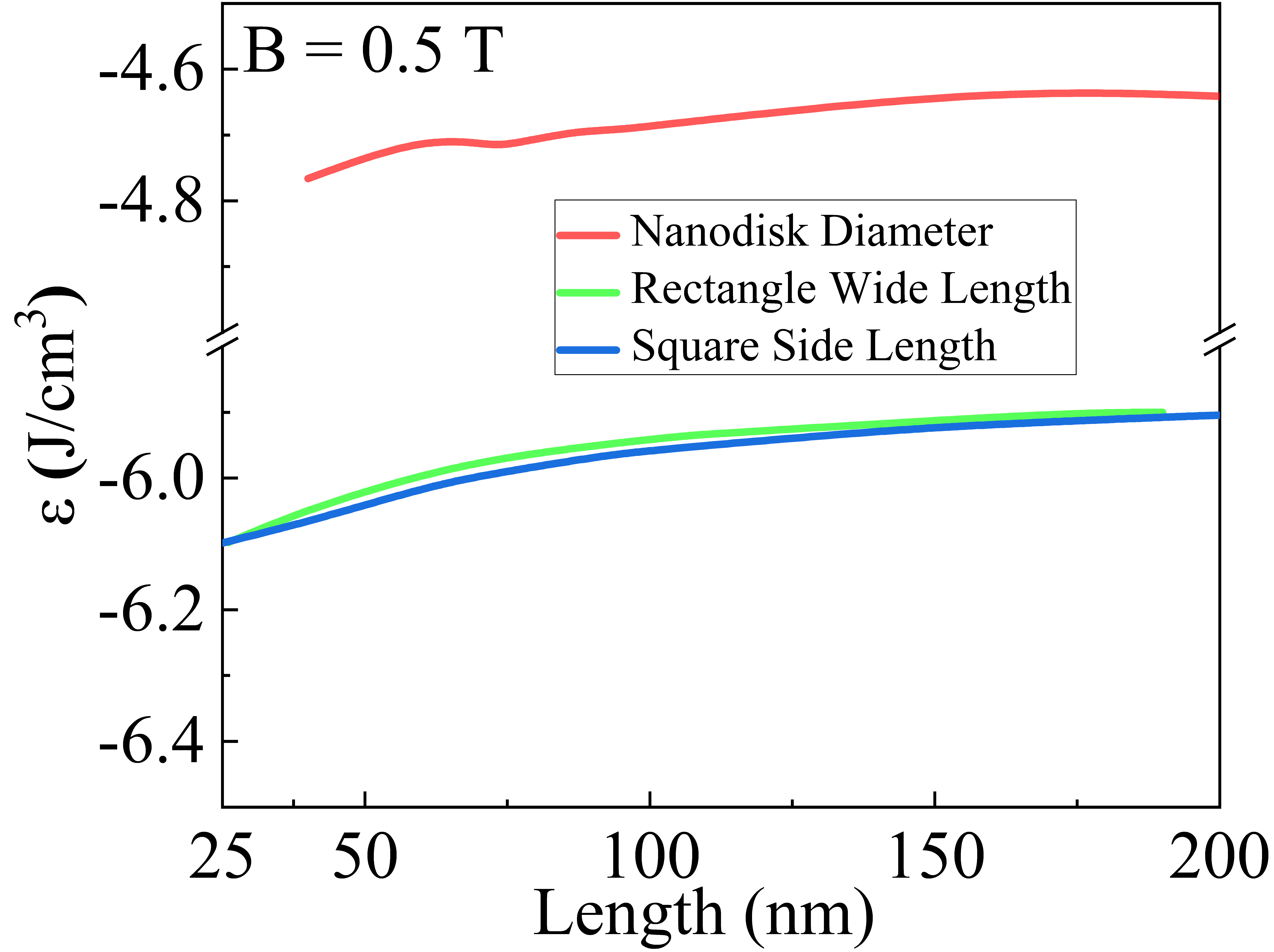}
\caption{The variation of the total magnetic energy density $\varepsilon_{tot}$ of the nanodisk, rectangle, and square as a function of size under a magnetic field of $B = 0.5 \, \text{T}$} \label{fig7}
\end{figure}
The simulation results depicted in Fig. \ref{fig7} highlight the variation of the total magnetic energy density $\varepsilon_{tot}$ of the nanodisk, rectangle, and square concerning size under a magnetic field of $B = 0.5 \, \text{T}$. The results show that the magnetic energy density of the nanodisk consistently exceeds that of the rectangle and square. Compared with results obtained in the absence of a magnetic field, the magnetic field amplifies the disk's energy-barrier advantage through several mechanisms. The Zeeman energy $E_{\text{Zeem}}$ introduced by the magnetic field creates a competitive or synergistic relationship with other energy terms, which are exchange energy $E_{\text{ex}}$, DMI energy $E_{\text{DMI}}$, and anisotropy energy $E_{\text{ani}}$. The high symmetry of the nanodisk allows the magnetic energies to remain elevated even in a magnetic field.

The rotational symmetry of the disk allows for the stable existence of skyrmions or vortex states, even in the presence of a magnetic field. These non-collinear structures need to maintain a high exchange energy $E_{\text{ex}}$ and iDMI energy $E_{\text{DMI}}$. Although the demagnetization energy $E_{\text{deg}}$ of the disk is lower than that of rectangular or square shapes, its high $E_{\text{DMI}}$ and anisotropy energy $E_{\text{ani}}$ dominate the total energy density. The high magnetic energy density of the nanodisk under a magnetic field arises from the synergistic enhancement of its geometric symmetry across multiple energy terms. The magnetic field also has an inhibitory effect on energy competition within the rectangle or square. This characteristic suggests that the disk is more suitable as a carrier of high-stability topological states under a magnetic field. At the same time, the rectangle or square is more appropriate for low-power magnetic storage devices due to its energy-efficiency advantages. This discovery provides a crucial theoretical basis for the energy optimization and functional design of magnetic nanodevices.

\section{Conclusion} 
This study elucidates how geometric symmetry and size govern the stability of skyrmions and phase transitions in magnetic nanostructures. Nanodisks, with their rotational symmetry, support diverse topological states, such as skyrmions, skyrmioniums, and multi-state configurations, through a synergistic interplay among DMI, exchange, and demagnetization terms. In contrast, square and rectangular nanostructures are limited to simpler topological configurations by corner-induced demagnetization fields and symmetry-breaking effects. Perpendicular magnetic fields further modulate these states, enabling reversible transitions and expanding operational windows, particularly in rectangles where the aspect ratio enhances skyrmion stability. These results provide a quantitative framework for designing skyrmion-based devices, emphasizing the importance of precision in nanofabrication and geometric optimization.

\section*{Acknowledgments}
This work is funded by the Science and Technology Program of Xuzhou (KC25001), and funded by the National Natural Science Foundation of China (Grant No. 12374079), and by the Fundamental Research Program of Shanxi Province, China (No. 202103021224250), the Science and Technology Innovation Project of Colleges and Universities of Shanxi Province of China (No. 2020L0242), the Higher Education Reform and Innovation Project of Shanxi Province (No. J20230617).




\end{document}